# Context-Awareness for Service Oriented Systems*

Hatim Hafiddi, Hicham Baidouri, Mahmoud Nassar and Abdelaziz Kriouile

Mobile and Embedded Information Systems Lab., Models and Systems Engineering Team,
National Higher School for Computer Science and Systems Analysis,
Mohammed V Souissi University,
BP 713, Agdal Rabat, Morocco

**Abstract**
Today, service oriented systems need to be enhanced to sense and react to user's context in order to provide a better user experience. To meet this requirement, *Context-Aware Services* (CAS) have emerged as an underling design and development paradigm for the development of context-aware systems. The fundamental challenges for such systems development are context-awareness management and service adaptation to the user's context. To cope with such requirements, we propose a well designed architecture, named *ACAS*, to support the development of *Context-Aware Service Oriented Systems* (CASOS). This architecture relies on a set of context-awareness and CAS specifications and metamodels to enhance a core service, in service oriented systems, to be context-aware. This enhancement is fulfilled by the *Aspect Adaptations Weaver* (A2W) which, based on the Aspect Paradigm (AP) concepts, considers the service's adaptations as aspects.

## 1. Introduction

Today, Service Oriented Architectures (SOA) are being widely deployed to improve information systems development and interoperability. Moreover, the increasing use of mobile devices and infrastructure has enabled users to access services from any location and at any time. The convergence of mobile technologies (i.e., in terms of mobile devices and telecommunication infrastructures) and software engineering paradigms (i.e., especially the service paradigm) has brought about a new generation of information systems, based on the *Context-Aware Service (CAS)* paradigm, known as *Context-Aware Service Oriented Systems* (CASOS). CAS driven development of service oriented systems enables them to be context-aware and consequently to provide users with customized and personalized behaviors depending on their contexts. For example, in an M-tourism system, a context-aware Restaurants Searching service provides users with suggestions depending on their locations, preferences and even the used device capabilities. Generally, this kind of information is called context.

The ambiguity of the context concept and the multiplicity of context situations to be considered make CAS hard to build. Moreover, traditional approaches for CAS development produce services whose business logics are tightly coupled with both of context management and adaptation logics. Consequently, the result of such approaches is usually complex services whose rate of evolution and reuse is much reduced. The aforementioned statements highlight the need of a development approach [13] and a well designed architecture for efficient CAS development. In this paper we propose a well-designed architecture, named ACAS, to support CAS development. The remainder of this paper is organized as follows. We first present a scenario that concerns an M-tourism system which will be used in subsequent sections as an illustrating example. Section 3 outlines the fundamental layers of the proposed architecture. In the following sections, we will outline the layers enabling the enhancement of core services to be context-aware. Section 8 briefly compares related works. In Section 9, we give a brief conclusion and outline our plans for future work.

## 2. Motivating Scenario

The following *Restaurants Searching* scenario illustrates the potential benefits of context-awareness for an M-tourism system:
*"Mr. Joseph, a French tourist, wants to taste the local gastronomy of Marrakech which he is visiting for the first time. So he gets connected via his mobile device (e.g., PDA, iPhone, BlackBerry, etc.) to a context-aware M-tourism system in order to obtain a list of restaurants that may meet his needs. After logging in, he makes a request. The system then proposes an adequate list of restaurants (restaurants availability is taken into consideration), close to his location (taking into consideration the GPS localization), described in his language (the system will consider the user's language) and taking account of his*

\* This work is supported by the FNSRSDT under the CSPT-ICTESAD project

*preferences (e.g., food preferences, restaurants prices, etc.). Also, let's note that such a system will resort, if necessary, to a results pagination mechanism to improve the responsiveness of the system (considering the device capacities, the RAM capacity and processor power in this case) and in case it detects any change in the tourist's context (e.g., weak battery or switching of connection mode from a high mode to a low one), it will automatically adapt its behavior (e.g., returned restaurants information will not include photos) for purposes of optimization (i.e., reducing latency and saving battery)."*

This scenario illustrates that CASOS systems differ from traditional systems since they use sensed information to adapt their services to the current user context. To that end, this class of systems is supposed to:
- Sense and compose context information from different sensors;
- Autonomously detect relevant changes in the context in order to dynamically adapt their services;
- Interoperate with third-party service providers (e.g., weather provider).

## 3. ACAS Architecture

CAS based development of context-aware systems involves several challenges. For instance, context definition (e.g., which context information is relevant for the adaptation of the application) and acquisition (e.g., collection from either native or web sensors) is not an evident process. In addition, the adaptation process must be based on mechanisms, in accordance with the best practices of software engineering (e.g., separation of concerns), to build well-designed CAS. Figure 1 illustrates the proposed architecture to tackle the fundamental challenges of CAS development. Through this architecture, our main objective is to enable CAS designers and developers to treat, while separating the concerns, the different activities related to the enhancement of core services (i.e., Services Layer) to meet context-awareness requirements. The proposed architecture is composed of the following layers:
- *Services Layer*: contains core services that fulfill the system business requirements;
- *Context Management Layer*: aims to deal with the main context management tasks such as context specification, representation and acquisition;
- *Adaptation Artifacts Layer*: provides the key concepts, necessary for core services adaptation, such as Adaptation Condition (i.e., situation involving services adaptation), Adaptation Rule (i.e., how to perform adaptations), Adaptation, etc.;
- *Context-Aware Services Layer*: specifies the variability of core services according to their use contexts. The core service and its variability form the Context-Aware Service;
- *Context-Awareness Layer*: providing Context Management and Adaptation Artifacts Layers is not sufficient to adapt core services to the context. The Context-Awareness Layer aims to provide a set of services that enable the adaptation of core services to the context in a rather abstract way (i.e., loosely coupling between core services and context-specific aspects through this layer).

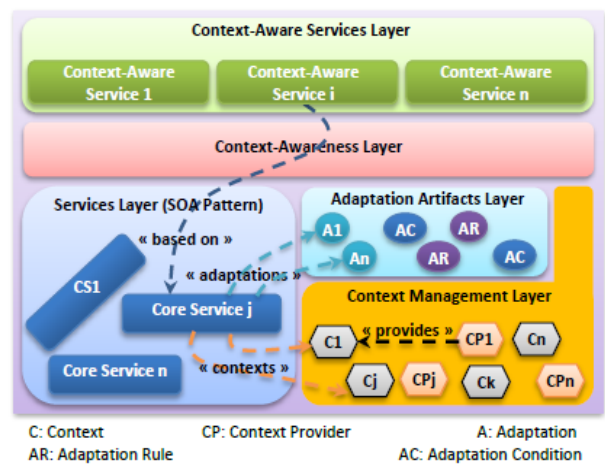

Fig. 1 ACAS architecture.

In the following sections, we will outline the layers enabling the enhancement of core services to be context-aware.

## 4. Context Management Layer

4.1 Context

Context is the information that characterizes the interactions between humans, applications, and the environment [5]. Context information is domain specific, as a type of information might be considered as context information in one domain but not in another (e.g., weather may be considered as a context parameter in a travel planning system but not in a money exchange one). Several context definitions serving various domains were proposed in the literature (e.g., [6], [25], [21], etc.). However the context definition given by Dey and Abowd remains the most generic. In fact, these authors defined context as *"any information that can be used to characterize the situation*

*of an entity. An entity is a person, place or object that is considered relevant to the interaction between a user and an application, including the user and applications themselves"* [8]. As given in [30], we consider context parameters as any additional information that can be used to improve the behavior of a service in a situation. Without such information, the service should be operable as normal; but with context information, it is arguable that the service can operate better or more appropriately [31].

Rather than giving a context formalization, case of figure for several researches on this topic, sometimes domain specific and sometimes generic but not very extensible, we choose to propose a meta-model [12] which is, at the same time, generic and abstract (Fig. 2). So, in this specification (see Fig. 1) a context is a set of parameters (e.g., language, localization, battery, connection mode, etc.) and entities (e.g., user, device, etc.) that can be structured on sub contexts. Sub contexts can also be recursively decomposed into categories. Context may be constituted of simple parameters (e.g., language), derived parameters (i.e., computed from other parameters; for example a distance parameter can be computed from two GPS positions) and complex parameters (e.g., GPS) which have representations (e.g., DMS (Degrees, Minutes, and Seconds) and DD (Decimal, Degrees) representation for the localization parameter).

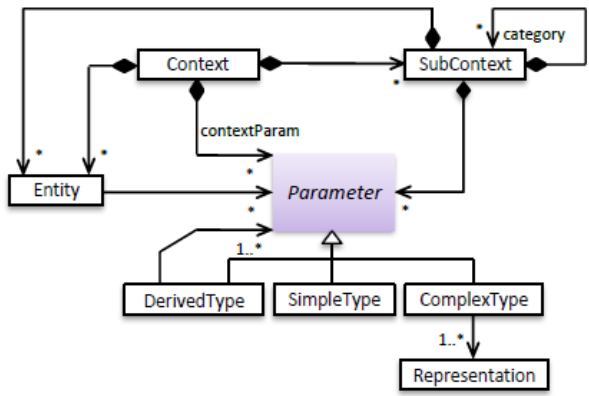

Fig. 2 Context metamodel.

To illustrate our metamodel, let's project it on the case of figure of the E-tourism system presented in the second section. The context for this system is composed mainly of the following sub contexts (see Fig. 3):
- *DeviceSubContext*: it contains parameters that describe the entity *Device*. It breaks up into two categories which are the software category (e.g., operating system, navigator type, supported type of data, etc.) and the hardware category (e.g., processor type, battery level, memory size, etc);
- *UserSubContext*: it is a sub context that contains parameters describing the entity *User* (e.g., preferences, localization, profile, etc);
- *EnvironmentSubContext*: this sub context contains the *Environment* parameters (e.g., time, weather, etc).

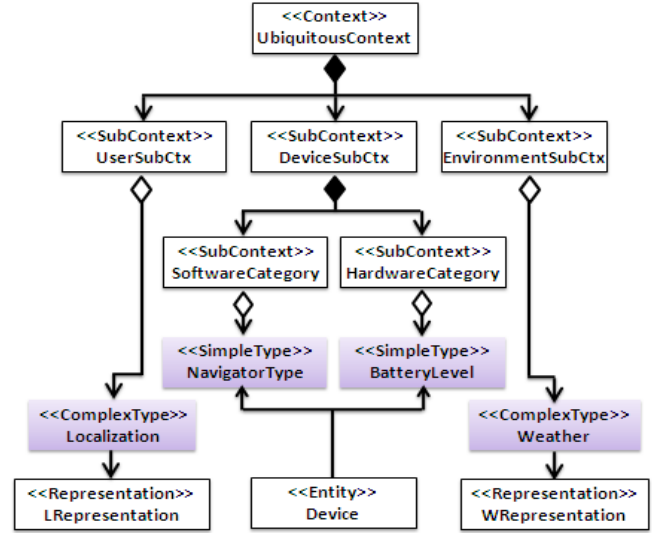

Fig. 3 Succinct context model for the M-tourism scenario.

### 4.2 Context Providers

The role of context providers is to gather context information from different sources such as sensors, web services, databases, etc. The process of collecting context information depends on the nature and the sources of context parameters. For instance, the user profile information is explicitly provided by the user and so it is characterized by an infrequent change. However, context parameters collected from sensors are subject to frequent changes. Their collection requires interaction with distributed and heterogeneous software or hardware sensors.

To abstract CAS developers from sensors and sensed data variety and complexity, we provide a context provider specification. In our specification, as illustrated in figure 4, a context provider (i.e., collector of a given service execution context) aggregates a set of parameters or entities providers. Both of these may dispose of an interface that specifies whether the provider is remote (e.g., a web service that provides weather information) or local (e.g., GPS sensor in a mobile device) and what mode of requests is supported (i.e., query-based or notification-based). A provider may use or derive from other providers to get context information.

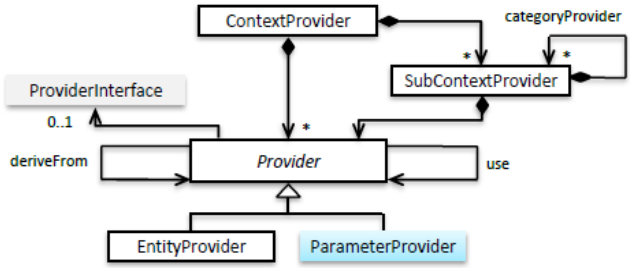

Fig. 4 Context provider metamodel.

Figure 5 shows the *Restaurants Searching* context provider composed of two EntityProviders: "DeviceProvider" providing context parameters that describe the device entity and "UserProvider" presenting context parameters that describes the user entity. It is also composed of two parameter providers: "TimeProvider" and "WeatherProvider". The latter has a provider interface that specifies its services and the supported mode.

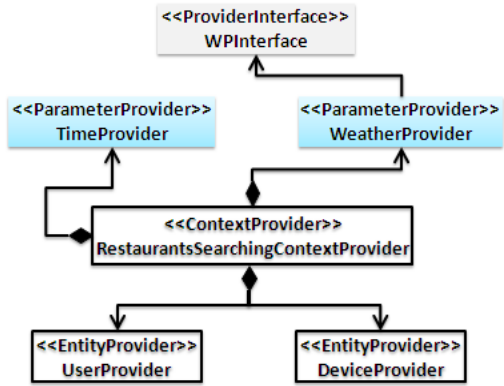

Fig. 5 Succinct *Restaurants Searching* context provider.

## 5. Adaptation Artifacts Layer

We introduce in this layer the concept of adaptation strategy (*AdaptationStrategy*) as an artifact used to specify the adaptation policy of a service to its current context of execution. So, an adaptation strategy (i.e., *SimpleAdaptationStrategy*) (Fig. 6) aggregates a set of artifacts indicating when (i.e., *AdaptationCondition*: classical condition expressed on context parameters) and how (i.e., *AdaptationRule*: defines the place in the service where the dynamic adaptations will be realized) a set of adaptations (i.e., *Adaptation*) must be applied, on the core service, in order to provide the expected behavior regarding the current execution context.

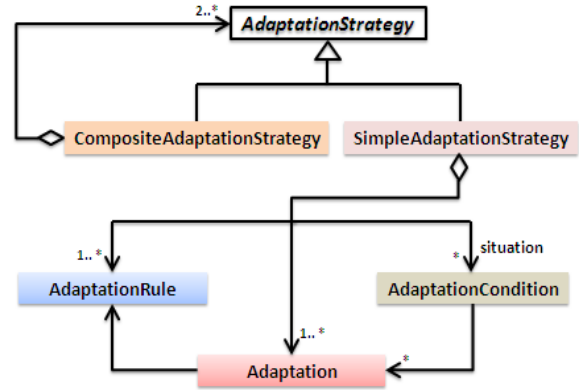

Fig. 6 Adaptation strategy metamodel.

## 6. Context-Aware Services Layer

One of the first uses of the term context-aware appeared in 1994 [24]. A service is context-aware if it provides customized and personalized behavior to users depending on their contexts [8]. To be context-aware, a service must be able to dynamically adapt its behavior to its several execution (i.e., use) contexts. In other words, the service (i.e., core service) must possess mechanisms so as to exploit only relevant information of the execution context and dynamically adapt its behavior. Henceforth, this appropriate context information relating to a specific execution situation forms what is termed the ContextView, and the result of the service adaptation to this *ContextView* forms the *ContextViewService*. Figure 7 presents the ContextView meta-model. Thus, a ContextView is seen as a set of context parameters that may aggregate other ContextViews.

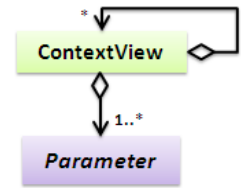

Fig. 7 ContextView metamodel.

The proposed CAS meta-model [13] is shown in figure 8. Accordingly, CAS is seen as a specific service with a number of ContextViews. For each, we associate an adaptation strategy (i.e., *CVSAdaptationStrategy*) that specifies the adaptation policy of the service to this ContextView. The adaptation result forms the *ContextViewService*. So, for a given service, the set of its ContextViewServices (CVSAdaptationStrategies) forms the CAS (*CASAdaptationStrategy*).

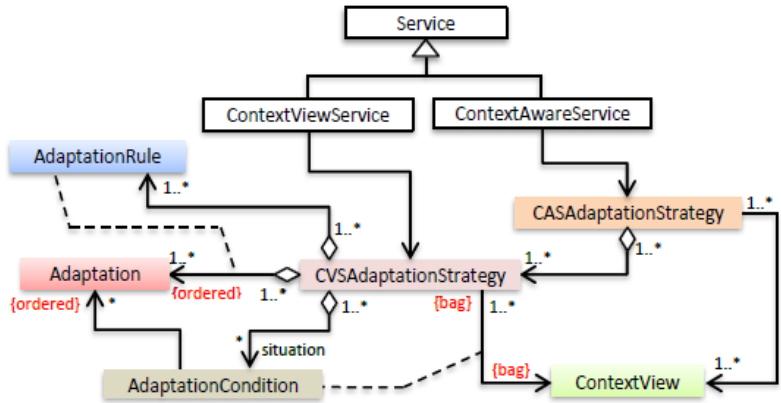

Fig. 8 CAS metamodel.

For instance, the *Restaurants Searching* Service has the following ContextViews:
- *User*: we associate to this ContextView the "UserAS" strategy. The latter consists in adapting the *Restaurants Searching* service to the tourist profile and his restaurants preferences;
- *Time*: we associate to this ContextView the "TimeAS" strategy. This strategy consists in filtering the restaurants response, based on time, to get only available restaurants;
- *Location*: we associate to this ContextView the "LocationAS" strategy. This strategy allows to resort only to restaurants that are close to the tourist's location;
- *BatteryState (ConnexionMode)*: we associate to this ContextView the "BatteryStateAS" (ConnexionModeAS) strategy. This strategy will provoke service adaptation by reducing the amount of data returned whenever the "batteryState" is low (the "connexionMode" is changed from a high to a low connectivity).

The figure below shows for instance the "BatteryStateAS" composition.

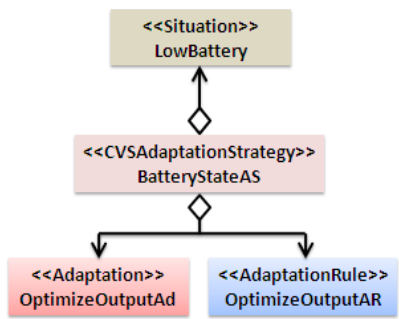

Fig. 9 BatteryStateAS adaptation strategy.

## 7. Context-Awareness Layer

### 7.1 Aspect Adaptations Weaver

The traditional approaches used for CAS design and development present several problems. In fact, simple core service duplication for each ContextView is a software engineering anti-pattern (e.g., high-cost of maintenance), also integrating adaptation logics into core service makes it complex and decreases its ability to be reused and maintained. Therefore, to rationalize the development and maintenance of CAS, we have to resort to new mechanisms and strategies that allow core service extension without any duplication or regression risks. Such mechanisms will favor loosely coupling between the core service and its adaptations seen as crosscutting concerns. CAS development can benefit from Aspect Paradigm (AP). AP [15] allows the modification of applications with so-called aspects. Aspects are modular units of functionality, used across the application code and woven at so-called pointcuts, which allow to transparently extend system functionalities. In our approach, the adaptations of a given service to its use contexts are seen as aspects. Thereby, the core service focuses only on business logic and all of its adaptations related to its ContextViews will be defined separately as aspects called *Adaptation Aspects*. These Adaptation Aspects will be dynamically woven at runtime into the core service, by our tool named *Adaptation Aspects Weaver (A2W)*, to produce the expected ContextViewService.

Figure 10 illustrates the mechanism behind the A2W tool. The *Request Notifier* notifies, in a synchronous or asynchronous mode, the *Decision Maker* with the executed service id and the execution context in order to recuperate the adequate CASAdaptationStrategy. Then, the Decision Maker inspects it in order to retrieve, based on context information availability, the current pertinent CVSAdaptationStrategy. The interpretation mechanism, operated by the *Service Reconfigurator*, consists in checking the AdaptationConditions to dynamically weave the required Adaptation Aspects, following a set of AdaptationRules, into the core service to produce the corresponding ContextViewService.

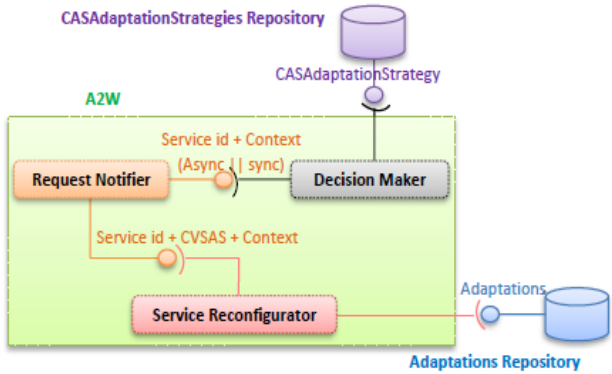

Fig. 10 A2W architecture.

As shown in figure 11, once the tourist has requested a proposition of restaurants, the *Restaurants Controller* (i.e., the entry of the system in a MVC pattern) gets the context of the executing service from the *Context Manager*, and then forwards the *request* with the recuperated *context* to the *Request Notifier*. This last notifies the *Decision Maker* with the appropriate *serviced* and *context*. Based on this information, the *Decision maker* retrieves the pertinent *CVSAdaptationStrategy* which will be used by the *Service Reconfigurer* in purpose to adapt the core service and provide a relevant response to the tourist expectations.

Figure 12 shows, for instance, two views for the restaurant searching service depending on the context state. Let us note that the user is a French tourist; in such a case, the screens are displayed using the French language. Once the user is authenticated, he has a set of services via the tab bar. For example, the user can consult a list of restaurants that suit his context through the "Restaurants" tab. Screen "a" shows the list of results in normal functioning conditions while Screen "b" shows the result in an optimization mode (without restaurants' pictures). This mode is activated, for instance, during a detection of a low "batteryState" or a low "connexionMode". The user can also sort results by relevance (i.e., pertinent restaurants), distance (i.e., distance between user and restaurant) or restaurant specialty. The result of the restaurants searching service is adapted based on context parameters such as time, device capabilities, weather, user profile, restaurants preferences and location.

## 7.2 Tools and Frameworks Support

The A2W tool is developed using Spring AOP framework [28]. The latter provides an API for the development of AOP concepts (e.g., aspect, advice, pointcut, etc.) and allows dynamic weaving. A2W plays the role of a mediator for dynamic adaptation of services to their execution contexts.

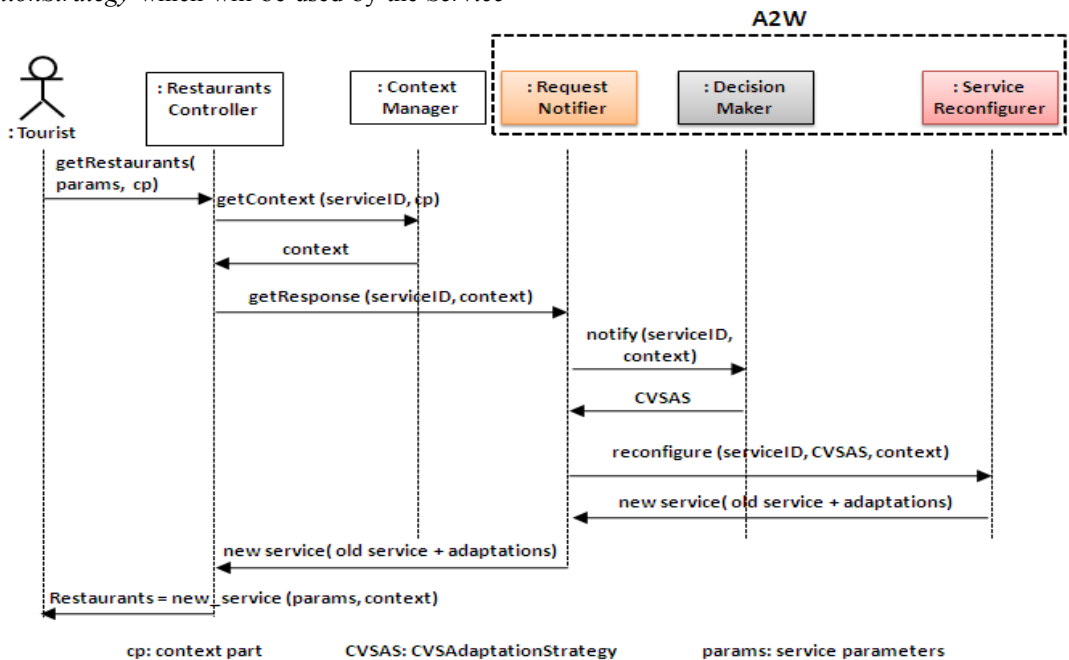

Fig. 11 Sequence diagram for the Restaurants Searching Service.

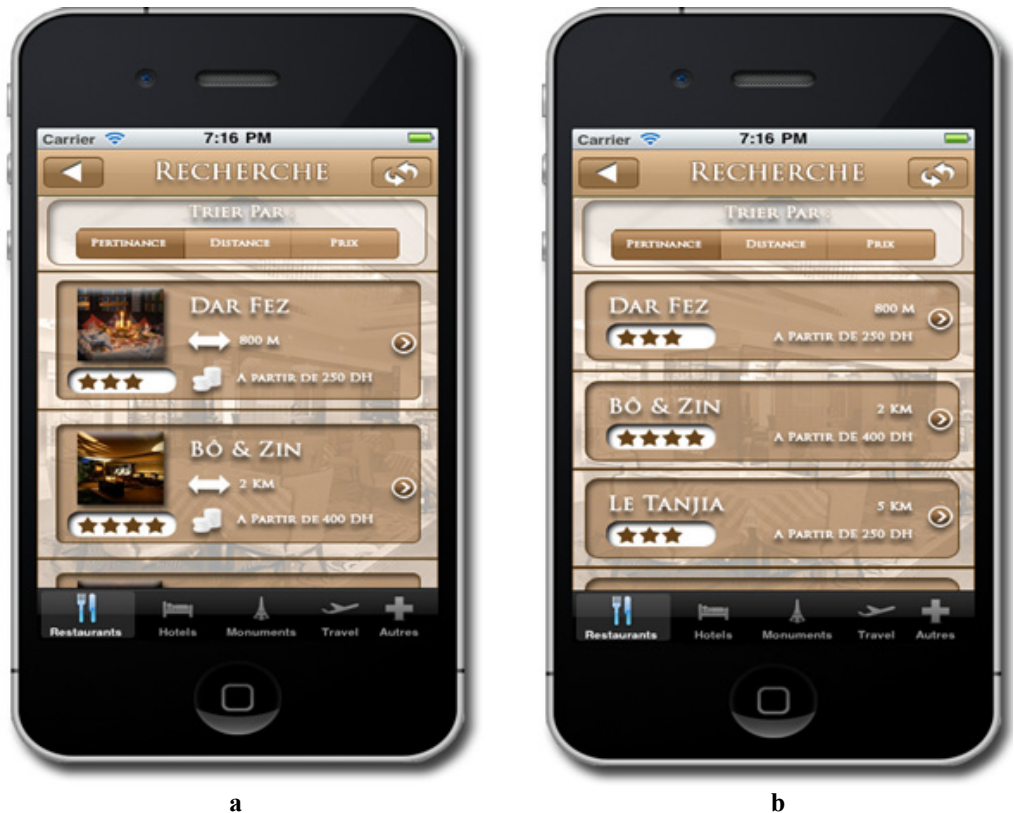

Fig. 12 Full and reduced views for the Restaurants Searching Service.

The client side supporting the motivating scenario is developed using the following development and deployment tools, and frameworks:

- *Development and Deployment tools:*
  - Eclipse EDI [10] for the development of the server side;
  - iOS SDK [3] for the development of the client side on iPhone and iPod devices;
  - Apache Tomcat 6 [1] integrated within the eclipse platform and used to deploy the server side;

- *Frameworks:*
  - Spring 2.5 [28] used as an IoC (Inversion of Control) container to link all the components of the system and also for transactions management;
  - Hibernate 3.3 [20] for persistence management;
  - CXF 2.2 [2] is the soap middleware that manage the communication between the client side and the server side using Web Services technology;
  - Configuration files (such as CASAdaptationStrategies) written using XML are parsed using JAXB2 OXM [18].

## 8. Related Work

As long as ACAS architecture combines a set of meta-models, a framework and a lightweight middleware for enabling context-awareness of services, we deal in this section with three categories of research works.

Several context models have been defined (e.g., Key-value pairs [23], databases (e.g., CML [14]), ontologies (e.g., CMF [17]), profiling (e.g., CC/PP [16]), etc.) and various context-aware middleware and frameworks have been developed (e.g., context Toolkit [22], CoBrA [7], K-Components [9], CORTEX [27], etc.) to handle context-aware applications development. In the one hand, the main objective of context modeling research works is to provide an abstraction of context information to permit easy context management. These research works do not deal with the adaptation of applications to the context. On the other hand, researches that focus on frameworks and middleware development try to simplify the development of context-aware applications by providing a set of services such as messaging, distribution, context management, etc. These research works do not deal with the modeling of context-awareness of applications and most of them suffer from the limited number of available context information, and the triggering of operations and context monitoring are defined statically at compilation time. So, due to the variety of context parameters to be collected and situations to be considered, we argue that context-awareness management needs the support of abstract context-awareness modeling.

In this context, some other works suggest the employment of model driven approaches for the development of context-aware applications. Authors in [29] define meta-models for modeling context-aware applications by planning several model views that model system context sensitivity, but they do not deal with adaptability. In our approach, service adaptability to the context is carried out through the CASAdaptationStrategy artifact and the A2W tool. Ayed [4] specifies an MDD (Model Driven Development) approach and a UML profile to design context-aware applications independently of the platform. He also proposes a design process that models the contexts that impact an application and its variability. The proposed approach does not deal with applications adaptation to the context. Grassi and Sindico provide support for context adaptation in [11] by decoupling the adaptation process from the application business logic. For this purpose they define a framework based on model-driven and aspect-oriented software development (AOSD). The proposed approach does not introduce the concept of entity in the context meta-model, and the underlying adaptation mechanism is not defined. In ContextUML project [26], Sheng and Benatallah define an approach for modeling context-aware Web Services. Context in ContextUML is specialized into "AtomicContext" and "CompositeContext", so the proposed meta-model does not refine context information. Moreover, authors do not specify the mechanism used to fulfill CAS adaptation. Authors in [32] focus on the context-aware development of web services oriented applications. They propose the use of model driven engineering and aspect oriented paradigm to separate concerns (i.e., business, context, context-awareness) in different models. The context meta-model proposed is domain specific and the use of AOP is limited to the composition of models but not for dynamic adaptation of services. Another important domain concerns Product Line Engineering (PLE) that has a great potential in modeling service variability. An important work is the one conducted in CAPPUCINE project [19]. Authors focus on context-aware adaptation in Dynamic Service-Oriented Product Line (DSOPL) rather than context modeling, and propose two different processes for the initial and iterative phases of product derivation. The main challenge to be faced in this work is to reduce non-deterministic behaviors when non deterministic context-aware assets are introduced. In our work, this challenge is faced by the execution of an ordered set of adaptations (i.e., priority management).

# 9. Conclusion

In this paper, we proposed an architecture for context-awareness of services named ACAS. The main purpose of this architecture is to cope with the fundamental challenges inherent in the enhancement of core services, in service oriented systems, to be context-aware. To make up for this limitation, we designed four layers: Context Management, Adaptation Artifacts, Context-Awareness and Context-Aware Services. To deal with these layers development, we proposed a set of meta-models. Thus, we presented a context meta-model which is generic and open to allow its extension to various domains depending on needs, and a context provider meta-model serving to abstract from the huge variety of context parameters and the complexity of context sensors. Then, we put forward a CAS meta-model and an adaptation mechanism, based on the Aspect Paradigm, which considers the adaptations of a service to its execution context as Adaptation Aspects dynamically woven by the A2W tool at runtime.

We focused in this paper on proposing a well-designed architecture to enable context-awareness of service oriented systems. We project to use the proposed meta-models for transformation purposes. So, both business and context-awareness models, in conformance with the proposed meta-models, can be transformed into platform specific models. The transformation process will rely on meta-models mapping and PIM (Platform Independent Model) to PSM (Platform Specific Model) transformation rules. We also plan to include our meta-models (context, context provider, CAS) in the Eclipse Modeling Framework (EMF), use the Graphical Modeling Framework (GMF) to build a graphical editor that will allow designers to model context management and CAS artifacts, and develop a transformation rules plugin to automate code generation.

**Hatim Hafiddi** received the Engineer of state degree in Software Engineering from National High School of Computer Science and Systems Analysis (ENSIAS) in 2007. He also received his PhD in Computer Science from the same School in 2012. His research interests are Context-Aware Service-Oriented Computing, Aspect Oriented Engineering, Mobile Information Systems Engineering, and Model-Driven Engineering.

**Hicham Baidouri** received the Engineer of state degree in Software Engineering from Mohammadia School of Engineers (EMI) in 2007. He is currently a PhD student in the IMS (Models and Systems Engineering) Team of SIME Laboratory at ENSIAS. His research interests are Context-Aware Service-Oriented Computing, Aspect Oriented Engineering, Mobile Information Systems Engineering, and Model-Driven Engineering.

**Mahmoud Nassar** is Professor and Head of the Software Engineering Department at National Higher School for Computer Science and Systems Analysis (ENSIAS), Rabat, Morocco. He is also Head of IMS (Models and Systems Engineering) Team of SIME Laboratory. He received his PhD in Computer Science from the INPT Institute of Toulouse, France. His research interests are Context-Aware Service-Oriented Computing, Component based Engineering, and Model-Driven Engineering. He leads numerous R&D projects related to the application of these domains in Embedded Systems, e-Health, and e-Tourism.

**Abdelaziz Kriouile** is a full Professor in the Software engineering Department and a member of SI2M Laboratory at National Higher School for Computer Science and Systems Analysis (ENSIAS), Rabat. He is also a Head of the SI3M Formation and Research Unit. His research interests include integration of viewpoints in Object-Oriented Analysis/Design, Service-Oriented Computing, and speech recognition by Markov models. He has directed several Ph.D thesis in the context of Franco-Moroccan collaborations.